\documentclass[final]{svjour3}
\usepackage[]{graphicx}
\usepackage{rotating}
\usepackage{amssymb}
\usepackage{cite}
\usepackage{easy-todo}
\usepackage{siunitx}
\usepackage{authblk}

\makeatletter
\journalname{Journal of Low Temperature Physics}
\titlerunning{Low-energy tails in x-ray absorbers}
\authorrunning{O'Neil, et al.}
\newcommand {\urss}[1]{\ensuremath{_{\mathrm{#1}}}}

\begin{document}

\title{On low-energy tail distortions in the detector response function of x-ray microcalorimeter spectrometers}

\author{G. C. O'Neil$^1$ \and P. Szypryt$^{1,2}$   \and E. Takacs$^{1,3}$ \and J. N. Tan$^1$ \and S. W. Buechele$^{1,3}$ \and A. S. Naing$^{1,4}$ \and Y. I. Joe$^{1,2}$ \and D. Swetz$^1$ \and D. R. Schmidt$^1$ \and W. B. Doriese$^1$ \and J. D. Gard$^2$ \and C. D. Reintsema$^1$ \and J. N. Ullom$^1$ \and J. S. Villarrubia$^1$ \and Yu. Ralchenko$^1$}

\institute{1. National Institute of Standards and Technology, Boulder, CO, 80303, USA, Tel.:303-497-4794,
\email{galen.oneil@nist.gov} \\ 2. University of Colorado, Boulder, CO, 80303, USA \\ 3. Clemson University, Clemson, SC 29634, USA \\ 4. University of Delaware, Newark, DE 19716, USA}

\maketitle

\begin{abstract}
We use narrow spectral lines from the x-ray spectra of various highly charged ions to measure low-energy tail-like deviations from a Gaussian response function in a microcalorimter x-ray spectrometer with Au absorbers at energies from 650~eV to 3320~eV. We review the literature on low energy tails in other microcalorimter x-ray spectrometers and present a model that explains all the reviewed tail fraction measurements. In this model a low energy tail arises from the combination of electron escape and energy trapping associated with Bi x-ray absorbers.

\keywords{transition edge sensor, low-energy tail, point spread function, microcalorimeter, JMONSEL}

\end{abstract}

\section{Introduction}
Arrays of transition-edge sensor (TES) microcalorimeter x-ray spectrometers have a combination of high collection efficiency and high resolving power that enable many otherwise difficult or impossible experiments. For example, they have been used for table-top time-resolved x-ray absorption and emission spectroscopy\cite{ONeil2017, Miaja-Avila2016}, probing the strong force with hadronic atoms\cite{Tatsuno2016a}, and partial fluorescence yield x-ray near-edge absorption spectroscopy of dilute samples\cite{Titus2017a}. The spectrometers used in these and other experiments had TES arrays with pixels consisting of a Mo-Cu bilayer with an absorber made from evaporated Bi\cite{Doriese2017}.  In each of these examples, the analysis is complicated, and the signal-to-noise ratio is reduced by the presence of a low-energy tail in the detector response function. Thus, we are motivated to study the low-energy tail to enable better analysis of x-ray spectra acquired with a TES spectrometer, and to guide the design of TES pixels with better detector response functions.

This low-energy tail is often modeled as an exponential with two parameters: the tail fraction $f\urss{tail}$ and the length scale $l\urss{tail}$ such that the detector response function at energy $E$ to monochromatic x-ray radiation with energy $E_0$ is

\begin{equation}
    DR(E) = [(1-f\urss{tail})\delta(E-E_0)+f\urss{tail} H(E_0-E) l\urss{tail}^{-1} e^{-(E_0-E)l\urss{tail}^{-1}}]*G(E-E_0,\sigma)
\label{eq:tailshape}
\end{equation}
where $H$ is the Heaviside function, $\delta$ is the Dirac delta function, $G$ is a normalized Gaussian function with standard deviation $\sigma$, and $*$ represents convolution. In the case of $f\urss{tail}=0$ this reduces to the ideal case of a Gaussian detector response function. In this paper we focus on low-energy tails with length scale on order 10~eV, which have been reported in the literature across many types of x-ray absorbing materials and energies with tail fractions from $\sim$0\% to 28\%. We do not discuss further the tails due to electron escape\cite{Porter1997} that occur on much larger energy scales.

The detector response function is distorted when  some fraction of the incident x-ray energy is not thermalized on the timescale of the TES response (typically $\sim1$~ms). Mechanisms that cause non-thermalized energy include escaping photo-electrons, escaping characteristic radiation, and energy being trapped in long-lived metastable states and released over a longer timescale. The large low-energy tail fractions observed in evaporated Bi absorbers are attributed to energy trapping associated with the physics of Bi and possibly associated with grain boundaries\cite{Yan2017}. In this paper we will refer to ``Bi energy trapping" without specifying the mechanism. A comparison of TES spectrometers with absorbers made from evaporated Bi, electroplated Bi, and Au found unmeasurably small tails in both the Au and electroplated Bi\cite{Yan2017}. However, those measurements could not probe tail fractions below $\sim5$\% because they were made using the characteristic x-ray radiation of transition metals, which have natural linewidths comparable to the tail length scale. A study using wavelength dispersive x-ray optics to generate x-ray radiation with $\sim1$~eV bandwidth found tail fractions from 6\% at 850~eV to 2\% at 8050~eV in TES pixels with electroplated Bi absorbers\cite{Eckart2019}. 

\section{The Array}

\begin{figure}[htbp]
\begin{center}
\includegraphics[width=1.0\linewidth, keepaspectratio]{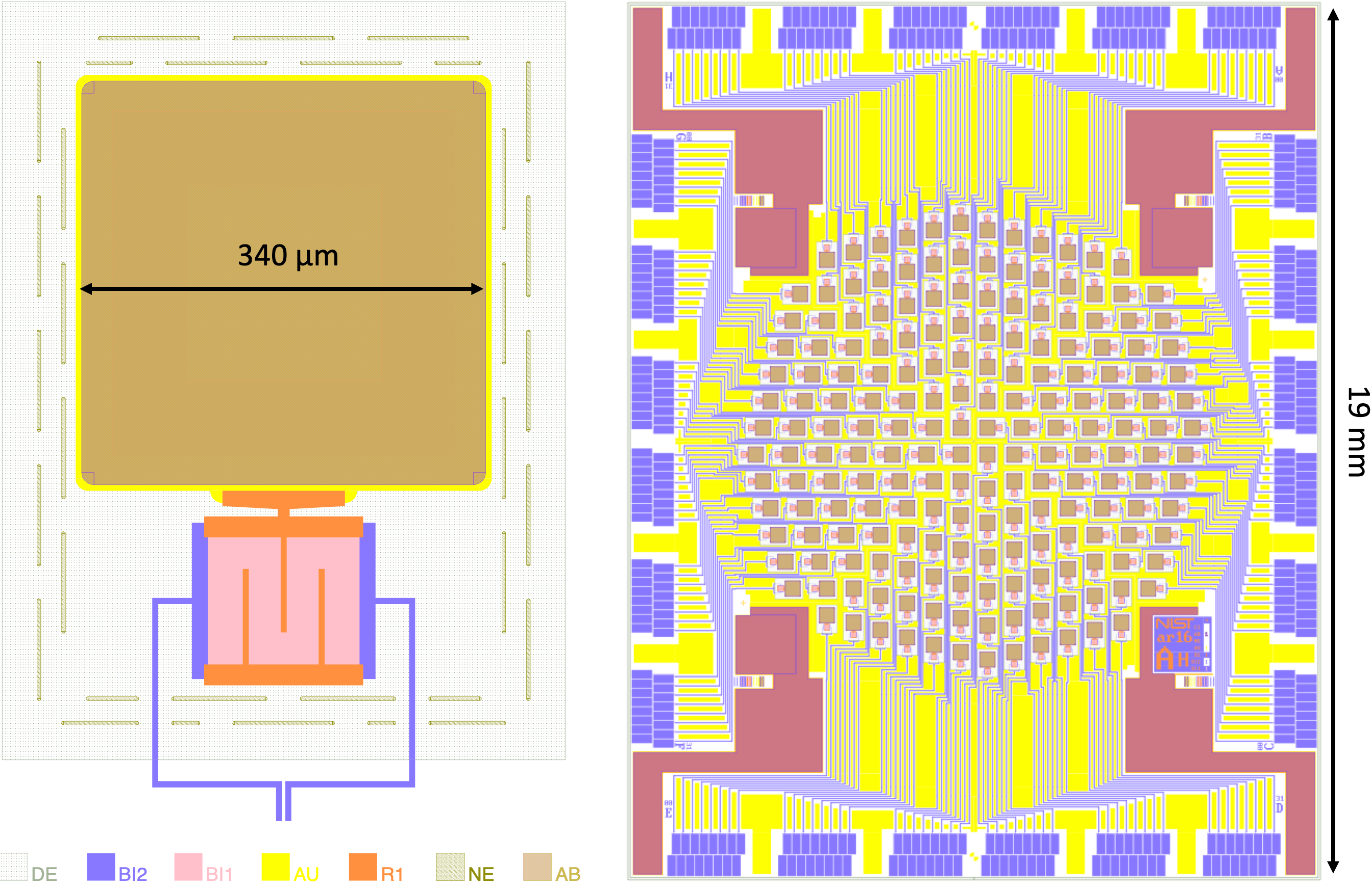}

\end{center}
\caption{(color online) \emph{left} Layout for the TES pixels used in this work with layer key at the bottom. \emph{right} The array layout, sharing the same color key. The layers are DE) deep etch to relieve the SiN membrane, BI2) 65~nm Mo used in the Mo-Cu bilayer for the TES as well as wiring and bond-pads, BI1) 215~nm Cu in the Mo-Cu bilayer,   AU) 186~nm thick Au layer used for heatsinking across the array as well as the x-ray absorber, R1) 419~nm Cu used for normal metal features on the TES and and thermal link to the Au absorber, NE) SiN membrane perforation etch to control thermal conductivity from the TES to the bath, and AB) an additional 779~nm Au layer for the absorber. \label{fig:pixel}}

\end{figure}

We measured the detector response function in an array of 192 TES pixels with Au sidecar absorbers. We expected very small low-energy tail fractions due to the lack of Bi and therefore the lack of Bi energy trapping. The pixel design and array layout are shown in in Fig. \ref{fig:pixel}. The absorber \emph{sidecar} is adjacent to the superconducting bilayer element of the TES, and is a \SI{1}{\um} thick Au square with side length \SI{340}{\um}. 

An aperture chip was installed on top of the array to minimize x-rays strikes at locations other than the Au absorber. The aperture chip has an array of \SI{280}{\um} square openings for each pixel. The Si thickness is \SI{220}{\um} and it has an additional \SI{0.5}{\um} Au layer. 

\begin{figure}[htbp]
\begin{center}
\includegraphics[width=0.48\linewidth, keepaspectratio]{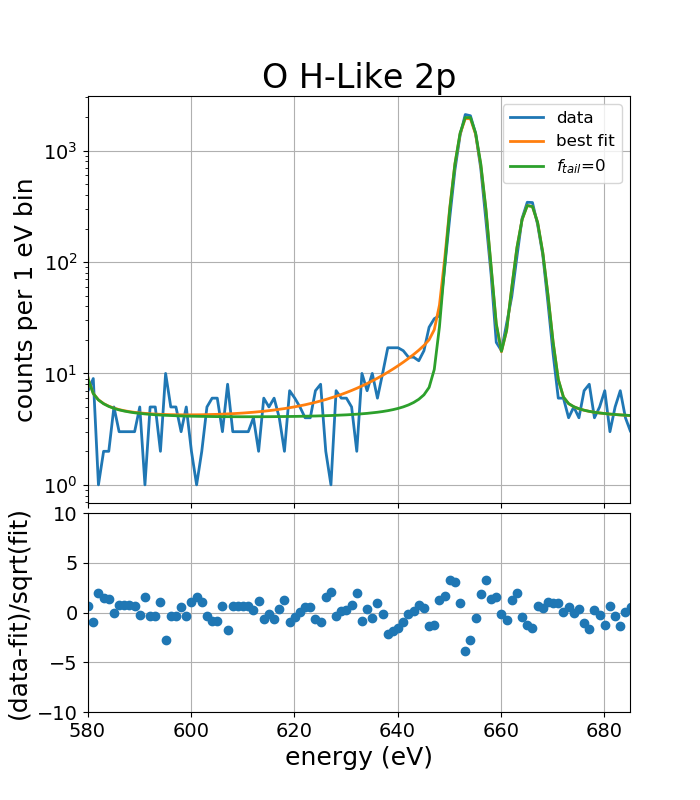}
\includegraphics[width=0.48\linewidth, keepaspectratio]{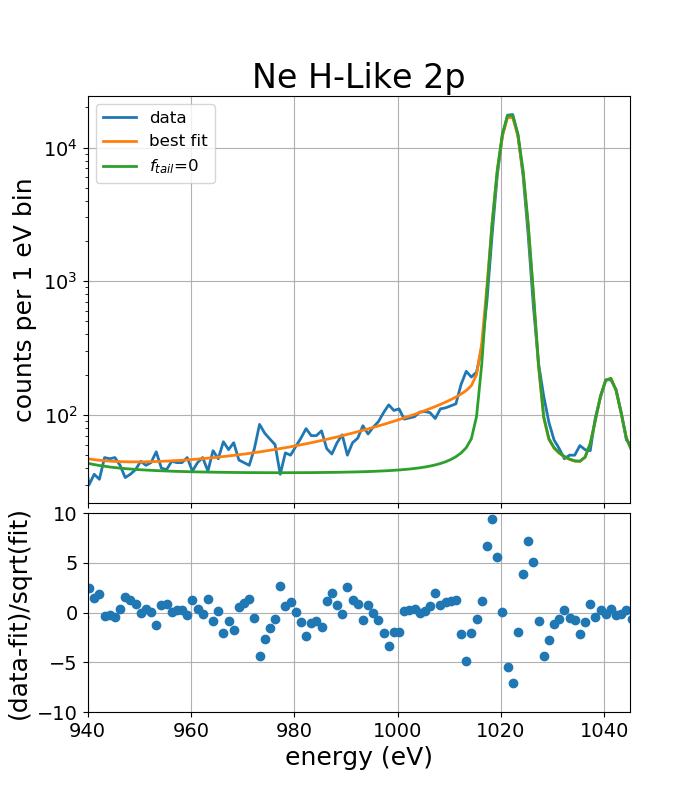}
\includegraphics[width=0.48\linewidth, keepaspectratio]{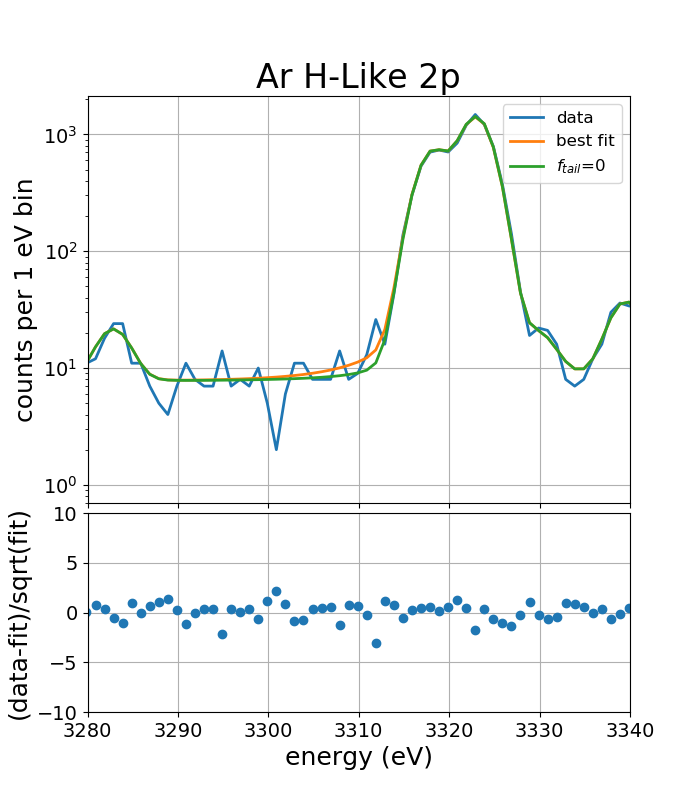}

\caption{(color online) Three sub-figures each containing data and fits from one emission line, named in the title, used to measure the low-energy tail parameters at a particular energy. In each case, two fits are visible; the best fit is used to extract the tail parameters and an additional fit performed with $f\urss{tail}$ fixed at 0 is shown as well. The best fit value of $f\urss{tail}$ was 0.028$\pm$0.004 for the O H-Like 2p line, 0.038$\pm$0.004 for the N H-Like 2p line, and 0.01$\pm$0.01 for the Ar H-Like 2p line. The difference between these two fits is the upper limit of the contribution to the spectrum that can be attributed to a low-energy tail. The lower segment of each sub-figure shows the residual of the best fit scaled by the square root of the best fit. The data are co-added from 97 detectors. The fits include nearby visible spectral lines to model the background including lines not shown that are within 20~eV of the plotted range. \label{fig:fits}}
\end{center}

\end{figure}

\begin{figure}
\begin{center}
\includegraphics[trim={0 5cm 0 0 },clip,width=0.96\linewidth, keepaspectratio]{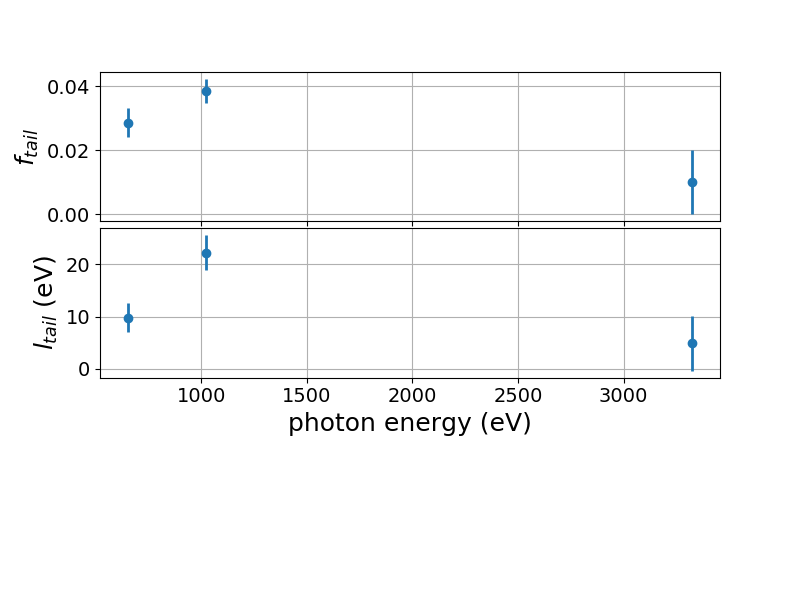}

\caption{(color online) \emph{top} Tail fraction $f\urss{tail}$ measured at three energies. The error bars are determined by the fitting routine. \emph{bottom} Tail length scale $l\urss{tail}$ measured at 3 energies. The tail fraction at 3320~eV is consistent with zero. If the tail fraction is zero, the measured tail length would not be a meaningful parameter. \label{fig:tail_vs_energy}}
\end{center}
\end{figure}

\section{Low-energy tail measurements}

A TES spectrometer otherwise similar to those described in Ref. \cite{Doriese2017}, but with the array described here, has been installed at the NIST Electron Beam Ion Trap (EBIT). An EBIT is a powerful tool for studying highly charged ions for applications related to fusion power, astrophysics and testing the frontiers of quantum electrodynamics\cite{Gillaspy2001}. Here we take advantage of the relatively narrow spectral lines available in the x-ray spectra of highly charged ions to study the low-energy tail in the TES spectrometer. During the commissioning of this spectrometer, we took x-ray spectra of H-like and He-like O, Ne, and Ar. In this naming scheme, highly charged ions are referred to as being like the neutral atom with the same number of electrons, for example O$^{7+}$ is H-like O. We examined these spectra to identify spectral features well suited to measuring the low-energy tail. We looked for spectral features with peak-to-background ratios exceeding 100 and relatively simple backgrounds $\sim50$~eV below these strong lines. 

We analyzed the spectra of the $2p\rightarrow1s$ transition in H-like O, Ne and Ar. We used the NIST Atomic Spectra Database\cite{Kramida} to find the energies of the two lines that make up this transition, and made reasonable guesses as to the intensity ratio. We modeled these lines with Lorentzian lineshapes with widths of 0.1~eV (much less than the detector resolution). We generated a fitting function by convolution of the Lorentzian lineshapes with the detector response function $DR$ and added Lorentzian components to account for weaker features observed near these lines, as well as a constant background level. We allowed both $f\urss{tail}$ and $l\urss{tail}$ as well as the intensities and locations of the various lines and the constant background level to be determined by fits to the data. Fits were performed with the Python package \emph{lmfit} with a residual function modified to maximize the poisson likeliness, rather than minimize the sum of squares of residuals. Minimizing the sum of squares is known to introduce biases into fits of histograms with few counts per bin\cite{Fowler2014}.  Figure \ref{fig:fits} shows these fits, as well as fits performed with $f\urss{tail}=0$ to allow visual assessment of the magnitude of the tail. We analyzed spectra coadded from 97 pixels, so these measurements represent an average behavior across the array, whereas the measurements in Refs. \cite{Eckart2019, Yan2017} were performed on single pixels. 

The tail fraction is found to be near 0.03 for energies 650~eV and 1020~eV, and consistent with zero at 3320~eV, as shown in Fig. \ref{fig:tail_vs_energy}. 

\section{Discussion}

\begin{figure}[]
\begin{center}
\includegraphics[width=0.7\linewidth, keepaspectratio]{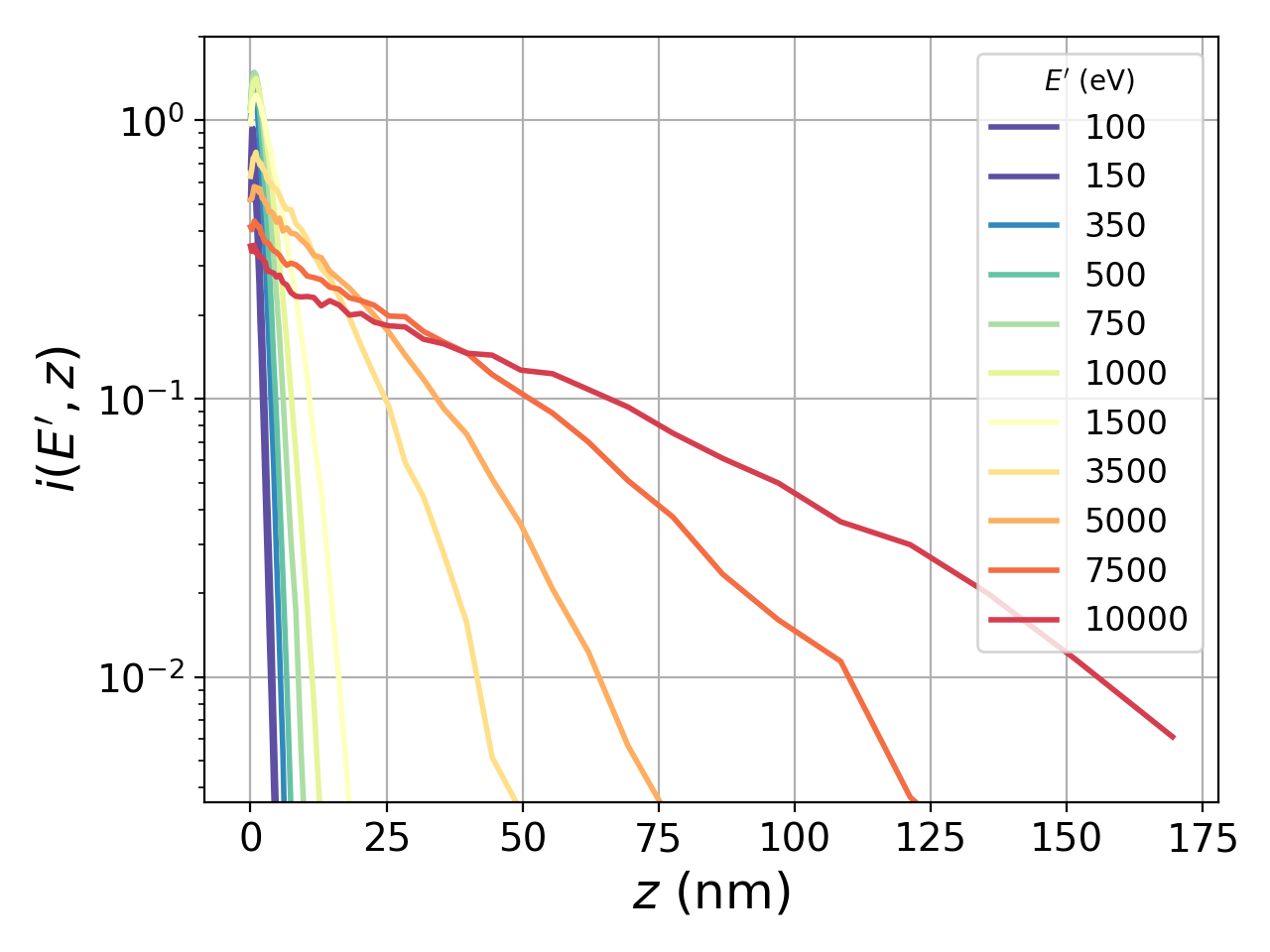}

\caption{(color online) The number of electron with energy less than 50~eV that leave the surface of the absorber when an electron of energy $E'$ is created at an absorption depth $z$, versus absorption depth. \label{fig:iez}}
\end{center}

\end{figure}

\begin{figure}[]
\begin{center}
\includegraphics[width=0.7\linewidth, keepaspectratio]{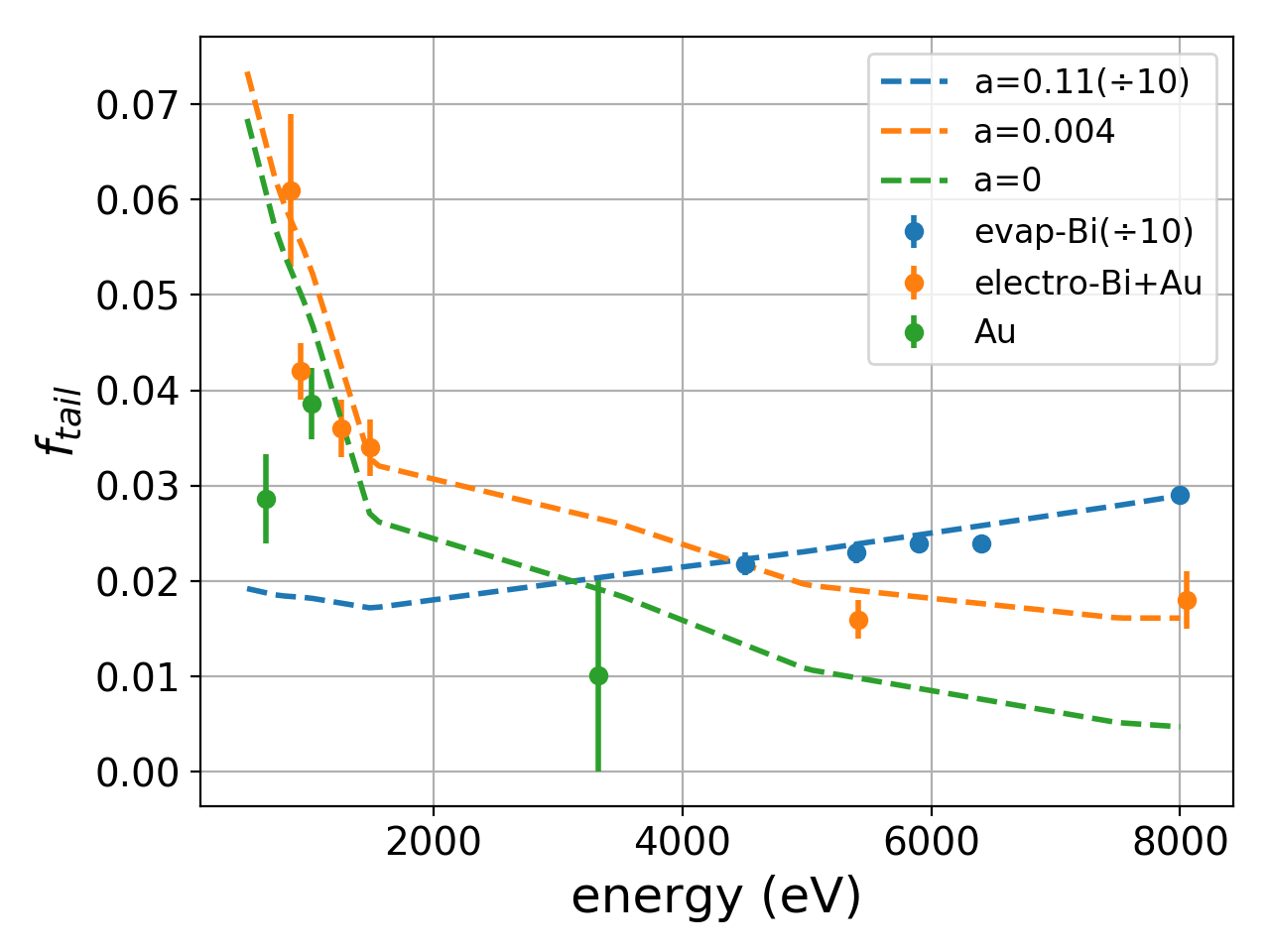}

\caption{(color online) Dots show tail fraction vs measured vs x-ray energy for Au absorbers (this work), evaporated Bi absorbers\cite{Yan2017} and electroplated Bi absorbers\cite{Eckart2019}. Lines show Eq. \ref{eq:ftail} evaluated for three different values of $a$ (shown in the legend) chosen to match each of the three absorbers. Note that the blue dots (evap-Bi data) and blue line ($a=0.11$ case) have been divided by 10 to allow us to compress the vertical axis.\label{fig:tail_frac_summary}}
\end{center}
\end{figure}

Here we consider the physical origin of the low-energy tails in Au absorbers in the previous section, as well as the low-energy tails in evaporated Bi absorbers\cite{Yan2017} and electroplated Bi absorbers\cite{Eckart2019} reported in the literature. We model the tail as arising from the combination of two effects: In the Au absorbers the tail is primarily due to electron escape whereas in the evaporated Bi absorbers the tail is due primarily to Bi energy trapping. The electroplated Bi absorbers show both effects. The tail fraction as a function of energy is modeled as 
\begin{equation}
f\urss{tail}(E) = f\urss{escape}(E)+f\urss{Bi}(E).
\label{eq:ftail}
\end{equation}
Tails due to Bi energy trapping are modelled empirically as linear function of $E$,
\begin{equation}
f\urss{Bi}(E) = a(1+b E), 
\end{equation}
where $b=1.9\cdot10^{-4}$~eV$^{-1}$ is determined by a fit to the evaporated Bi data, and $a$ is determined separately for each absorber. A linear form is chosen for $f\urss{Bi}(E)$ because it models the evaporated Bi data reasonably well. Tails due to electron escape are modeled with
\begin{equation}
f\urss{escape}(E) = \int_{0}^{t}{\frac{e^{-z/l(E)}}{l(E)} i(E',z) dz}.
\label{eq:fescape}
\end{equation}
where $E'$ is the energy of the photo-electron generated by an x-ray with energy $E$, $i(E',z)$ is the number of electrons with energy less than 50~eV that leave the surface of the detector when an electron of energy $E'$ is created at an absorption depth $z$, and $\frac{e^{-z/l(E)}}{l(E)}dz$ is the probability that an incident photon of energy $E$ is absorbed in between depths $z$ and $z+dz$, $l(E)$ is the absorption length in Au, and is calculated with the \emph{xraylib} Python package, and $t=$\SI{1}{\um} is the thickness of the Au absorbers.

We calculate $i(E',z)$, shown in Fig. \ref{fig:iez}, with a Monte Carlo electron transport simulation with code known as JMONSEL\cite{Villarrubia2015}. JMONSEL models electron scattering in atomic potentials, secondary electron generation, electron-phonon scattering, x-ray generation, and is optimized to track electrons with low energies. Many comparable codes cannot or do not normally track electrons with energies below 50~eV. JMONSEL does not model Auger electron generation. We modeled the absorber as a half-plane of Au, placed an ``electron gun" producing electrons with energy $E'$ with isotropic angular distribution at depth $z$ and a ``detector" counted the number $n$ of electrons with energy $<50$~eV that exited the half-plane. For each pair of $E'$ and $z$ we ran $N=20,000$ trials, and calculate $i(E',z)=n/N$. We used 11 values for $E'$ from 100~eV to 10,000~eV and 49 values of $z$ from 0~nm to 169~nm, each with logarithmic distributions. For $z$ values greater than 169~nm we set $i(E',z)=0$; inspection of Fig. \ref{fig:iez} shows that extending to absorption depths greater than 169~nm would begin to matter for energies of 7,500~eV and greater. We interpolated $i(E',z)$ along $E'$ as needed. We make a limited attempt to model the relationship between x-ray energy and photo-electron energy by making a simple approximation to account for the most significant absorption feature in the energy range of interest. For x-ray energies below the Au M$_5$ edge we use $E'=E$ and for energies above the the M$_5$ edge $E'=E-E_{M_5}$ where $E_{M_5}=2205$~eV is the Au M$_5$ edge energy. We note that the choice to subtract the M$_5$ edge from the x-ray energy has a small effect, it reduces $f\urss{escape}$ by a factor of roughly one quarter for energies near the edge and has nearly no effect for energy above 5,000~eV. 

Figure \ref{fig:tail_frac_summary} shows this model compared to our data on Au absorbers, as well as data on evaporated and electroplated Bi absorbers from Refs. \cite{Eckart2019, Yan2017}. With no free parameters in $f\urss{escape}$ and using $a=0$ to represent the lack of Bi energy trapping we reproduce all the tail fraction measurements in Au absorbers to within a factor of two in the worst case. The electroplated Bi absorbers have a 25~nm Au capping layer. Inspection of Fig. \ref{fig:iez} shows that, for energies below 3,500~eV, nearly all electron escape is due to x-ray absorption within 25~nm of the surface, so this model should work reasonably well for the electroplated Bi absorbers. Figure \ref{fig:tail_frac_summary} shows that the trends in the electroplated Bi absorber data are well explained by this model.  

Here we discuss potential steps to improve our understanding of low-energy tails and modify our detectors to reduce these tails. A natural extension of this model would be to more accurately model the relationship between x-ray energy and photo-electron energy, as well as to model Auger electron generation. JMONSEL can calculate energy spectra of the escape electrons, information that could be used to predict $l\urss{tail}$. We could use this model to predict the effectiveness of a low-Z capping material (e.g., Al or SiO) on top of a high-Z absorber at reducing tail fractions due to electron escape. It may be possible to reduce the portion of the tail due to electron escape with appropriate voltage biasing of the TES detectors and surrounding surfaces such that escape electrons are accelerated back to the absorber.

The work function of the Au absorber sets a minimum value of the energy removed by a single electron. One would expect that low-energy tailing caused by escaping electrons would have a distinctive shape as a result. We modified Eq. \ref{eq:tailshape} by replacing the Heaviside function by $H((E_0-\phi)-E)$ where $\phi=5$~eV is the work function of the Au absorber and by re-normalizing the exponential term. We found that this modification does produce a distinctive shape for the low energy tail. However, for each of the spectra shown in Fig. \ref{fig:fits} the difference between the best fit with and without this modification is far too small to be distinguished with our data.

\section{Conclusions}

We have used narrow spectral lines generated from various highly charged ions to measure the low-energy tail in the detector response function of x-ray microcalorimeter detectors with Au absorbers. Tail fractions were $0.04$ or lower for all energies measured. We reviewed the literature on tail fraction in microcalorimeter x-ray detectors, and suggest that the combination of escape electrons and Bi energy trapping provides a plausible explanation for all of the reviewed data. We provided a semi-empirical model that describes all of the reviewed data with moderate success; the electron escape portion of this model has no free parameters. Comparison with this model leads to the conclusion the tail fraction due to Bi energy trapping in the electroplated Bi absorbers described in Ref. \cite{Eckart2019} is 4\% of that in the evaporated Bi absorbers described in Ref. \cite{Yan2017}.

\begin{acknowledgements}
The authors thank Csilla Szabo-Foster for introducing the EBIT and TES teams to each other.
\end{acknowledgements}

\pagebreak

\bibliographystyle{ieeetr}
\bibliography{main}

\end{document}